%% file: main.tex
\documentclass[conference]{IEEEtran}

\usepackage{algorithm2e}
\usepackage[pdftex]{graphicx}
\usepackage{comment,authblk,flushend}
\usepackage{color}
\usepackage{caption}
\usepackage{subfigure}
\usepackage{url}
\usepackage{amsmath,amssymb}

\newcommand{\argmin}{\operatornamewithlimits{arg\ min}}

\newcommand\logten{\ensuremath{\log_{10}}}

\begin{document}
%
\title{Through-Wall Person Localization Using Transceivers in Motion}

%
\author[P. Hillyard et al.]
       {Peter Hillyard$^1$, Dustin Maas$^2$, Sriram Premnath$^{3}$,
       Neal Patwari$^{1,2}$, and Sneha Kasera$^4$
       \\
       $^1$Dept. of Electrical \& Computer~Engineering, University of Utah, Salt Lake City, USA\\
       Email: hillyard,npatwari@ece.utah.edu\\ 
       $^2$Xandem Technology, Salt Lake City, USA, Email: dustin@xandem.com\\
       $^3$Qualcomm Research, Santa Clara, USA, Email: sriramnp@qti.qualcomm.com\\
       $^4$School of Computing, University of Utah, Salt Lake City, USA, Email: kasera@cs.utah.edu\\
       }


\maketitle

\begin{abstract}
We develop novel methods for device-free localization (DFL) using transceivers \emph{in motion}.  Such localization technologies are useful in various cross-layer applications/protocols including those that are related to security situations where it is important to: know the presence and position of an unauthorized person; monitor the daily activities of elderly or special needs individuals; or gain situational-awareness in emergency situations when police or firefighters can use the locations of people inside of a building in order to save lives. We propose that transceivers mounted on autonomous vehicles could be both quickly deployed and kept moving to ``sweep'' an area for changes in the channel that would indicate the location of moving people and objects. The challenge is that changes to channel measurements are introduced both by changes in the environment and from motion of the transceivers.  In this paper, we demonstrate a method to detect human movement despite transceiver motion using ultra-wideband impulse radar (UWB-IR) transceivers.  The measurements reliably detect a person's presence on a link line despite small-scale fading.  We explore via multiple experiments the ability of mobile UWB-IR transceivers, moving outside of the walls of a room, to measure many lines crossing through the room and accurately locate a person inside within 0.25 m average error.
\end{abstract}

\input{Introduction}

\input{Methodology}
\input{RTI}
\input{ExperimentalVerification}
\input{RelatedWork}
\input{Conclusion}

\section*{Acknowledgements}
We would like to acknowledge the formative role that Dr.~Cliff Wang played in the generation of the idea to perform RTI using mobile transceivers.  This work is supported by the US Army Research Office under Grant \#W911NF-12-1-0361, and by the US NSF under Grant 0748206.


\bibliographystyle{IEEEtran}
\bibliography{overall}

\end{document}

%% file: Introduction.tex
\section{Introduction} \label{S:Intro}
Multistatic RF localization technologies, such as radio tomographic imaging \cite{wilson09a}, device-free passive localization \cite{youssef07}, multiple-input multiple-output (MIMO) radar \cite{haimovich08}, and multistatic ultra-wideband impulse radar (UWB-IR) \cite{chang04}, offer the potential to locate moving people and objects over wide areas using RF channel measurements. In highly cluttered multipath environments, these systems rely on the change in the RF channel to identify and locate moving people and objects.  Such localization technologies are useful in various cross-layer applications/protocols including those that are related to security situations where it is important to: know the presence and position of an unauthorized person; monitor the daily activities of elderly or special needs individuals; or gain situational-awareness in emergency situations when police or firefighters can use the locations of people inside of a building in order to save lives.

Device-free localization (DFL) (where people being localized do not carry any wireless transmitters) research has, typically, built algorithms and systems around the assumption that the position of transceivers performing DFL remain static, and that measured changes are solely due to movements in the environment \cite{kaltiokallio2014spatial,zhang2007tracking,chen2011montecarlo}.  Under these conditions, changes in channel measurements are due to the movement of people or objects in the environment \cite{wilson11fade,zhao2013radio}.  The location of moving people or objects can be accurately estimated based on which wireless links show significant changes in channel measurements.  The use of static transceivers poses certain practical challenges. First, a large number of transceivers may be required to cover the entire monitored area. Second, and very importantly, there may not be sufficient time to deploy a large number of transceivers when people must be localized very quickly (e.g., in emergency situations).

\begin{figure}[t!]
        \center{\includegraphics[width=3.5in]
        {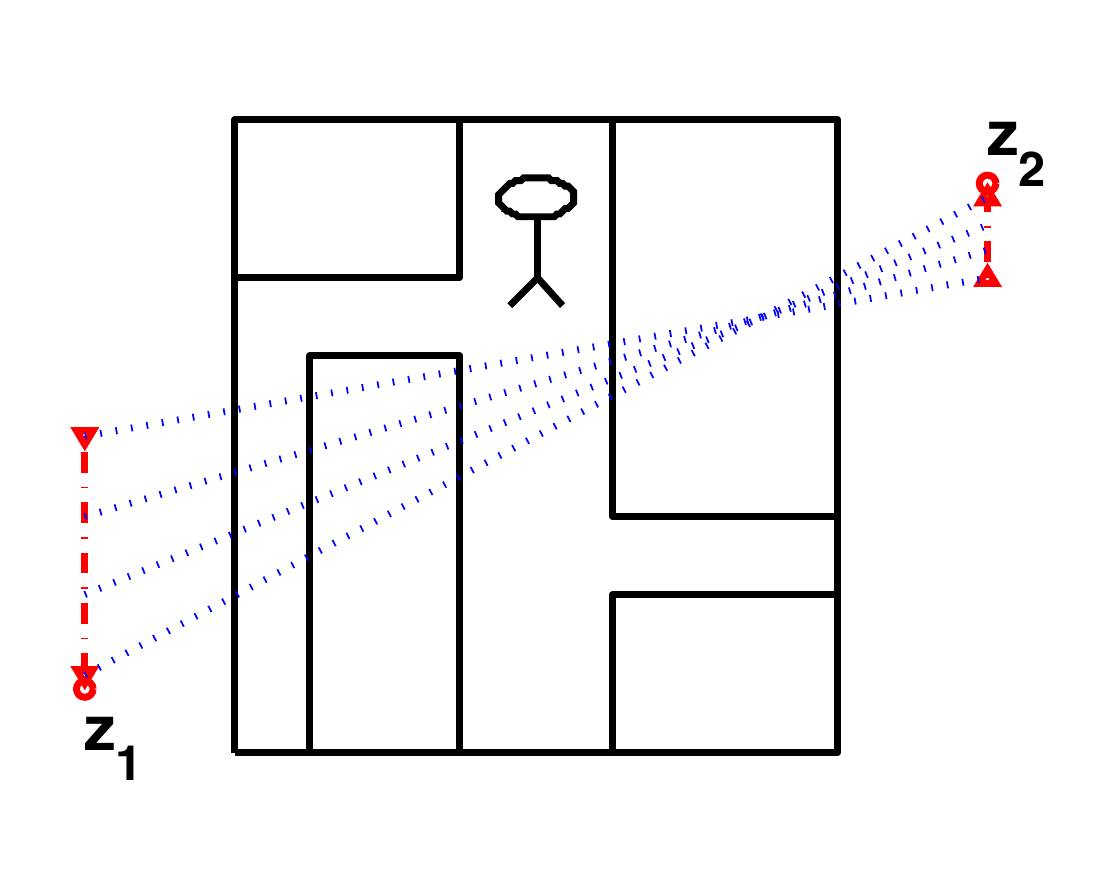}}
        \caption{In an example deployment, mobile transceivers autonomously move around the perimeter of a building while making measurements of the channel.  The person inside the building does not carry any wireless transmitters.  Mobile transceivers $\mathbf{z}_1$ and $\mathbf{z}_2$ autonomously move around the perimeter of the building.}
        \label{F:explanationFig}
\end{figure}

In this paper, we develop methods for DFL using only a few transceivers \emph{in motion}.  In our methods, the RF channel is measured by mobile transceivers (e.g., aerial or terrestrial vehicle-mounted) that can autonomously change position to enable rapid deployment, adapt to a moving target, or refine location estimates.  For example, pairs of mobile transceivers, like $\mathbf{z}_1$ and $\mathbf{z}_2$ in Figure~\ref{F:explanationFig}, can make rapid channel measurements while in motion then detect the obstruction of the \emph{link line}, (i.e., the line from the transmitter to receiver), an operation we call \emph{link line presence} detection.  By measuring link line presence on many moving link lines, the system can effectively sweep a building for activity. We assume that a transceiver can determine its coordinates relative to a fixed local coordinate system, and can eventually return to a previous position, to re-measure from that position.  The mobile node's self-positioning and self-navigation will not be perfect, and part of this work explores how accurate the self-positioning and self-navigation need to be in order to accurately localize people.  In a sophisticated deployment, a swarm of robotic vehicle-mounted transceivers may be used, some in motion and some static during any given period of time.

When mobile transceivers are used for measurements of a radio channel, multiple factors can cause variations in the channel measurements. First, with mobile transceivers, the radio channel between two transceivers changes due to small-scale multipath fading, a result of the change in position of the transceivers. Second, the radio channel also changes, by similar magnitudes, due to motion of a person nearby because of temporal fading.  Furthermore, a mobile transceiver does not necessarily measure the channel at fixed locations. Distinguishing the changes in the channel that result from alterations in the environment and those that result from location variations is a challenging task.  We tackle this challenge by examining the differences in the wireless channel multipath characteristics due to movement of people in a monitored region and due to small-scale fading caused by motion of transceivers monitoring the region. 

Existing research shows that fading rate variance is proportional to angular spread of a wireless signal~\cite{durgin2000}.  Angular spread can be shown to be very small in the first few nanoseconds of a channel impulse response (CIR) \cite{patwari1999peer,durgin2003wideband} presumably because for multipath to arrive close in time to the line of sight path, the paths must be contained within a narrow ellipse around the transmitter and receiver \cite{liberti1996geometrically}.  Thus, while small-scale fading will have dramatic effects on CIRs at medium and large time delays, we can expect it to have a minor effect at small time delays. 

In contrast, when a person or object moves across the link line, the person will scatter, absorb, and reflect the direct path and other paths with low excess delay.\footnote{We use the term direct path to mean the multipath that travels along the link link either in an unobstructed or obstructed manner.}  Specifically, the person will attenuate multipath with small time delays as a result of shadowing.  We distinguish the effects of small-scale fading from movement of transceivers from that of shadowing due to a person being near or on the line of sight between the transmitter and the receiver, by examining the first few nanoseconds of the impulse response measurement. 

We use the change in the first few nanoseconds of the measured impulse response, thus, to test for link line presence, even with transceivers in motion. The use of UWB-IR radios allow us to measure the multipath power in the first few nanoseconds of the measured impulse response. Quantifying motion on a link line with moving endpoints serves as a fundamental building block for \emph{environmental imaging} using networks of mobile transceivers for the applications described above.

In this paper, we make the following contributions. We first develop the components of our methodology for accurately localizing people, who are not carrying any transmitters, through walls using mobile transceivers.  This methodology applies radio tomographic imaging to measurements of energy in the first few nanoseconds of the measured channel impulse responses gathered by mobile transceivers in order to localize people within the monitored area.  We implement our methodology using two UWB transceivers and perform extensive experiments in different settings.  Through our measurement campaigns, we show that link line presence can be detected with great accuracy despite movement of the transceivers. Furthermore, we show that mobile transceivers can be used to localize a person to within 0.25 m, on average, of their actual position.

The use of mobile transceivers both complements \emph{and} sets our work apart from existing methods for multistatic UWB radar, which perform through-wall imaging using stationary transceivers \cite{chang04,paolini08}.  In contrast to traditional multistatic UWB radar, our paper provides methods that use measurements collected by \emph{mobile} UWB transceivers.  In addition, previous multistatic UWB research ignored the information contained in the direct path, and in fact, assumed that an intruder whose excess delay was very small (e.g., on the link line) could not be located \cite{paolini08}.  We show that the energy change at these low excess delays can be used to reliably detect link line presence, even with a mobile device.

This paper proceeds as follows.  Section \ref{S:Methods} describes our method of measuring the CIR; how we distinguish small-scale multipath fading from temporal fading; and we describe the framework by which we perform DFL.  In Section \ref{S:Experiments}, we perform experiments to first evaluate our methods for distinguishing small-scale multipath fading from temporal fading and second, to show that we can perform DFL with manually moved transceivers and accurately estimate a person's location.  We discuss existing research on localization, mobile devices, and link line presence detection methods in Section \ref{S:RelatedWork}, and we conclude this work and discuss future work in Section \ref{S:Conclusion}.

%% file: Methodology.tex
\section{Methodology} \label{S:Methods}

In this section, we describe our use of UWB-IR transceivers to measure the CIR.  In a through-wall experiment, we test and evaluate the feasibility of distinguishing small-scale multipath fading and link line presence.  We analyze the test to set conditions under which we can distinguish the two causes of variateion in channel measurements.  We end by presenting the framework by which we use mobile transceivers to perform through-wall DFL.

\subsection{Measured Impulse Response} \label{S:ImplementUWB}
The multipath channel causes multiple copies of the transmitted signal to be received, each copy with its own amplitude and propagation delay.  Specifically, the received signal $h(t)$ is written as \cite{maas2012channel,hashemi93},
\begin{equation} \label{E:CIR}
h(t) = \sum_{j=1}^{N} \alpha_j s(t - \tau_j)
\end{equation}  
where $N$ is the number of multipath, $\alpha_j$ and $\tau_j$ are the amplitude and propagation delay of the $j$th multipath, and $s(t)$ is the transmitted signal.  

Ideally, to measure the channel impulse response, we would make our transmitted signal $s(t)$ be equal to the Dirac impulse function $\delta(t)$.  This would allow the receiver to uniquely determine the amplitude of each component.  However, such a transmitted signal would consume infinite bandwidth.  The closest we can get in practice is to use UWB-IR.  

\begin{figure}[t!]
        \center{\includegraphics[width=3.5in]{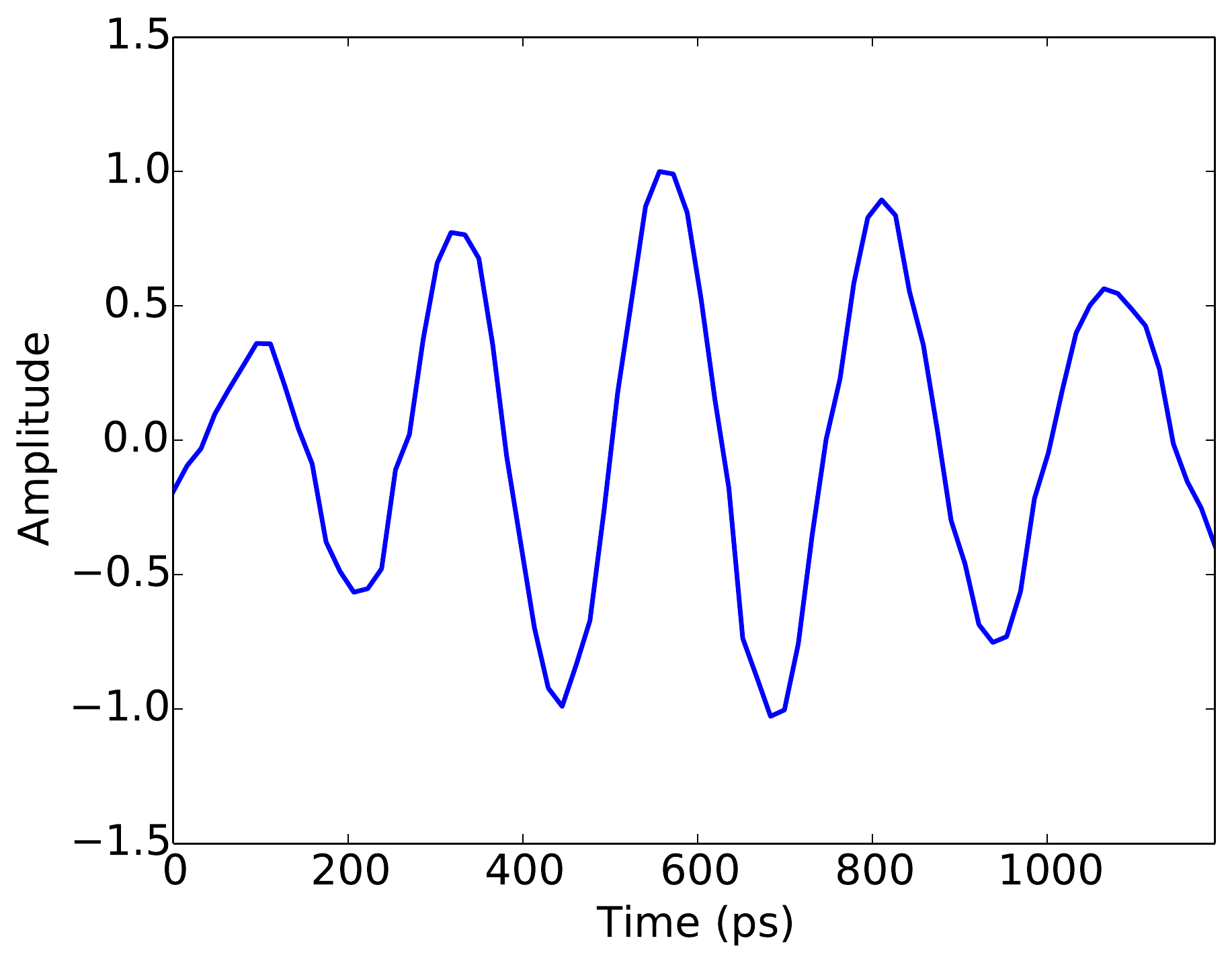}}
        \caption{\label{F:pulse} Measured Gaussian monocycle pulse $s(t)$ sent by our UWB-IR transmitter.}
\end{figure}

Our UWB-IR transmitter sends the Gaussian monocycle pulse $s(t)$ which is designed to be band limited to between 3 to 6 GHz; $s(t)$ closely resembles the pulse that we have shown in Figure \ref{F:pulse}.  The majority of the energy of the wide-band pulse occurs in a window whose duration is approximately 1 ns.  Paths that arrive within 1 ns of each other overlap and add together, either constructively or destructively.  One can see from the shape of $s(t)$ in Figure \ref{F:pulse} that two paths arriving 100 ps apart would tend to nearly cancel each other, while two paths 200 ps apart would add constructively.  Note that 100 ps translates into 0.03 m of path length, so even a small position change can result in the difference between constructive and destructive interference. 

However, multipath near the direct path contribute less small-scale multipath fading. Small-scale multipath fading occurs because spatial translation of an antenna changes the relative time delays of multipath \emph{at different rates}, thus bringing their sum in and out of destructive and constructive interference.  The rate of change of $\tau_j$ is a function of the angle of arrival of the $j$th path.  If two multipath arrive from the same angle, their time delays change at the same rate, and thus their sum does not change.  For paths arriving within a few nanoseconds of the direct path, the multipath must be contained in a very narrow ellipsoid with the transmitter and receiver locations as foci, and thus the angular spread of the arriving multipath is very low \cite{liberti1996geometrically}.  Thus we should see very slow small-scale multipath fading in the first few nanoseconds of the measured UWB received signal $h(t)$.

\begin{figure}[t!]
        \center{\includegraphics[width=3.5in]
        {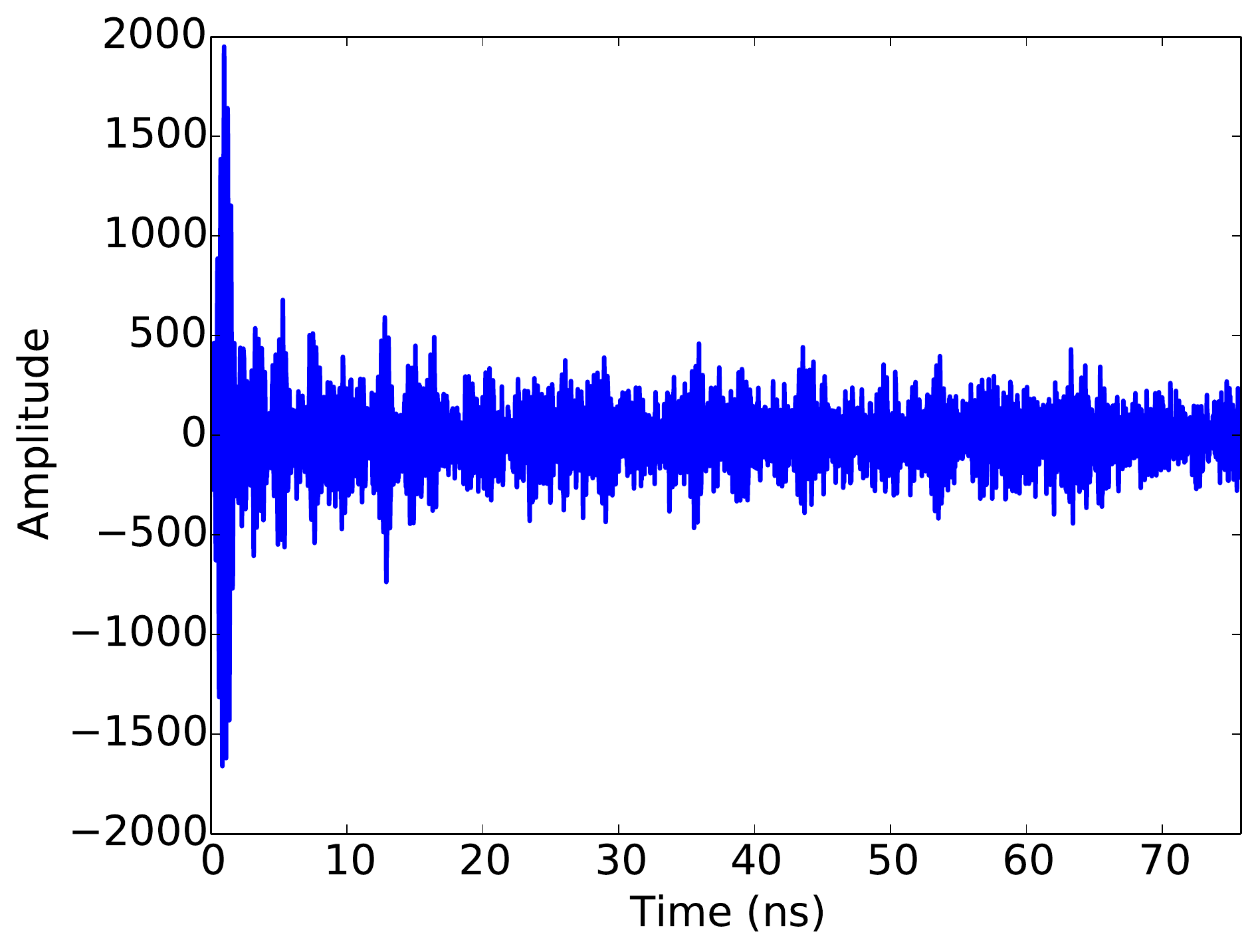}}
        \caption{\label{F:exampleCIR} An example of a measured impulse response $h(t)$ using UWB-IR.}
\end{figure}

In this paper, we use two UWB-IR transceivers (Time Domain, P220) \cite{timeDomain2013Online} with sampling period $T_s=15.89$ ps.  Figure \ref{F:exampleCIR} shows an example of the post-processed measured impulse response.  

The two transceivers are not time-synchronous (to the ps level), and thus processing is needed to time-align different measurements, using the first arriving multipath as time zero.  In this paper, we do this by cross-correlating the received signal $h(t)$ with the Gaussian monocycle pulse $s(t)$ and finding the first time at which the correlation coefficient exceeds a threshold $\rho$.  When $\rho$ is too small, we time-align with a later arriving multipath; when $\rho$ is too large, we time-align earlier than the first arriving multipath.  We use $\rho=0.75$ in this paper based on observations of time-aligning accuracy.  The occasional CIR measurements that were heavy corrupted with interfering signals were discarded.

A convenient way to represent the measured impulse response is using a power-delay profile (PDP).  The PDP shows the energy of the received signal as a function of excess delay $\tau$.  By doing so, we remove unnecessary details about the pulse shape $s(t)$.  The energy in time delay bin $m$, $e_m$, is calculated as,
\begin{equation} \label{E:PDP}
e_m = \int_{mT_w}^{(m+1)T_w} |h(t)|^2 \mathrm{d}\tau.
\end{equation}
The value of $e_m$ is the integral of the received power that falls in the $m$th $T_w$-wide window.  

\subsection{Proof-of-Concept Experiment} \label{S:thruWall_MEB3240_results}

Our intuition is that the first few nanoseconds of the PDP provides a means to be able to distinguish between small-scale multipath fading and link line presence.  To test this, we set up a simple through-wall experiment.  

\begin{figure}[t!]
        \center{\includegraphics[width=3.5in]
        {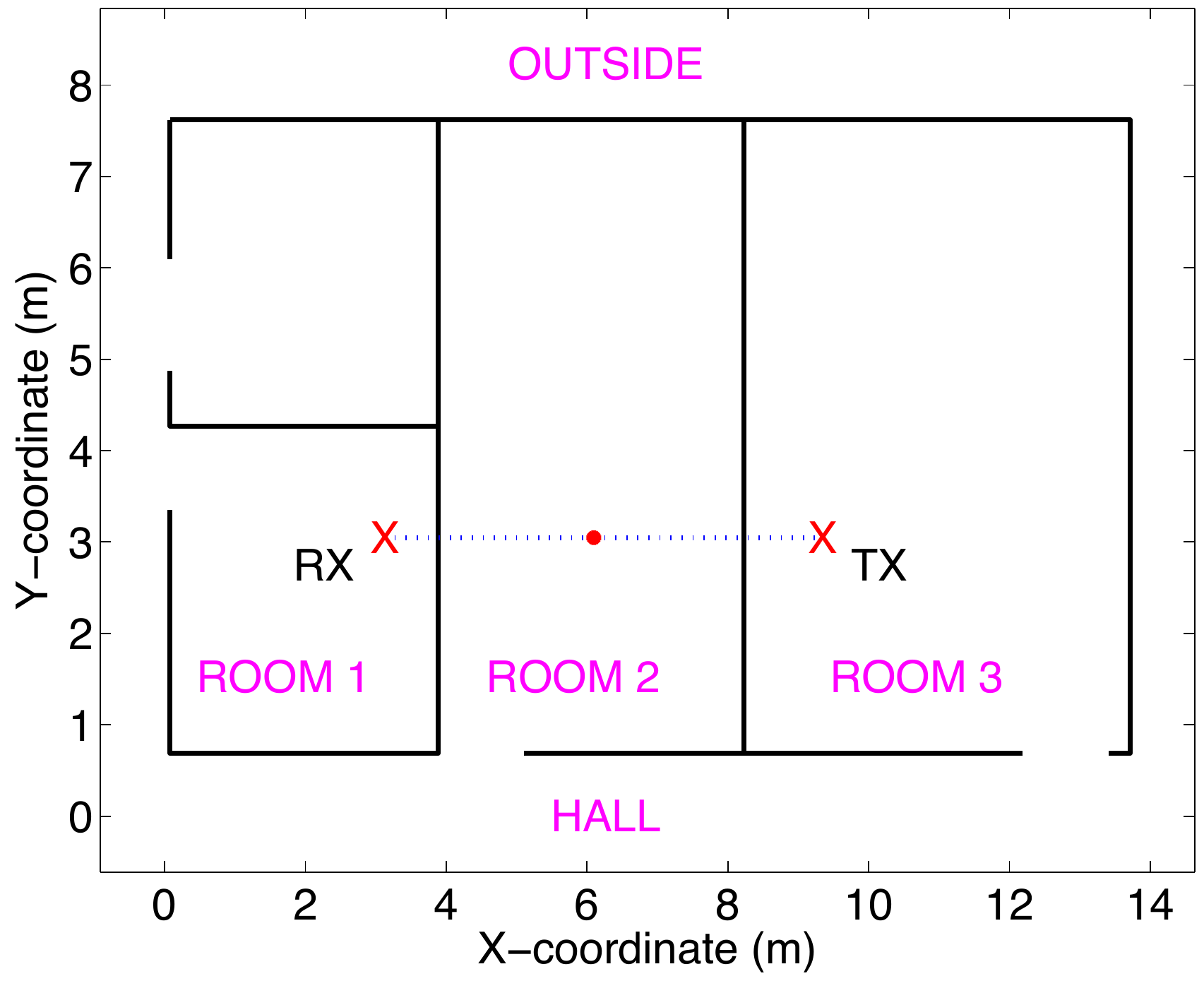}}
        \caption{\label{F:3240floorPlan} The floor plan showing UWB transceiver placement and the standing position of the person.  The red dot represents where the person stands while the measurements are taken.}
\end{figure}
\begin{figure}[t!]
        \center{\includegraphics[width=3in]
        {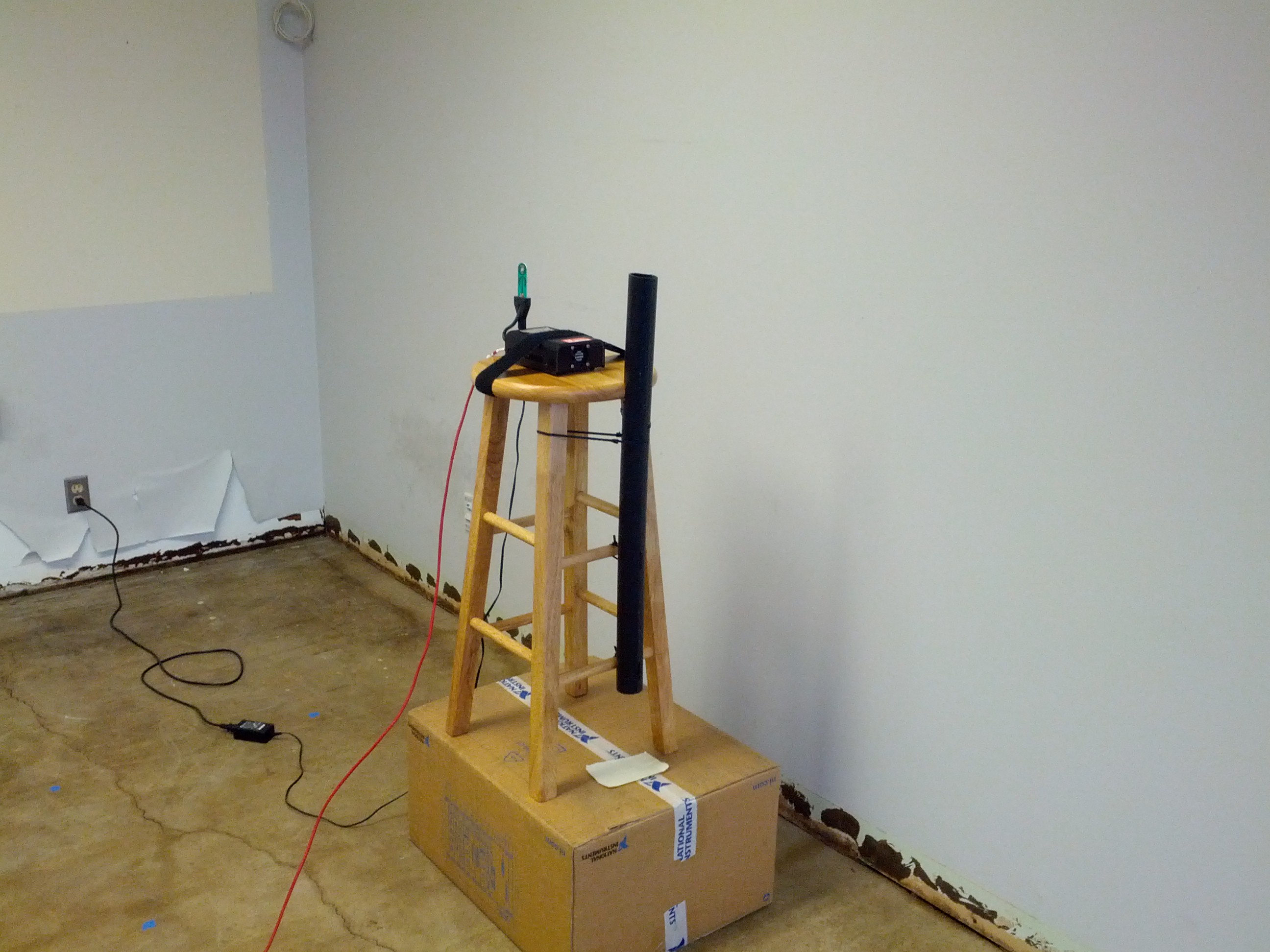}}
        \caption{\label{F:radioOnStool} One of the transceivers used to collect UWB impulse response measurements.}
\end{figure}

\begin{figure*}[ht!]
     \begin{center}
        \subfigure[]{%
            \label{F:rxMove&empty}
            \includegraphics[width=0.4\textwidth]{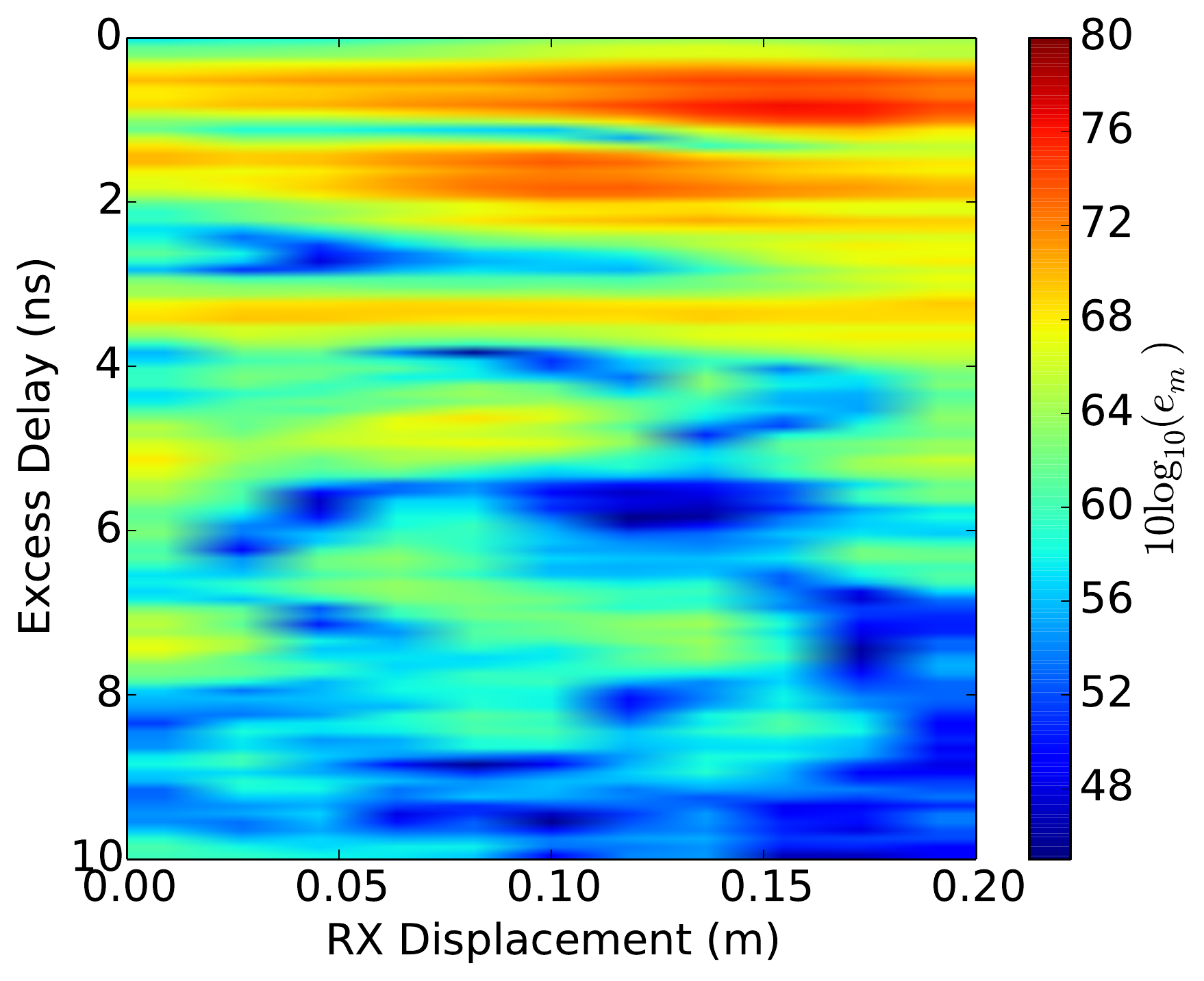}}%
        \quad\quad\quad\subfigure[]{%
          \label{F:rxMove&onLOS}
          \includegraphics[width=0.4\textwidth]{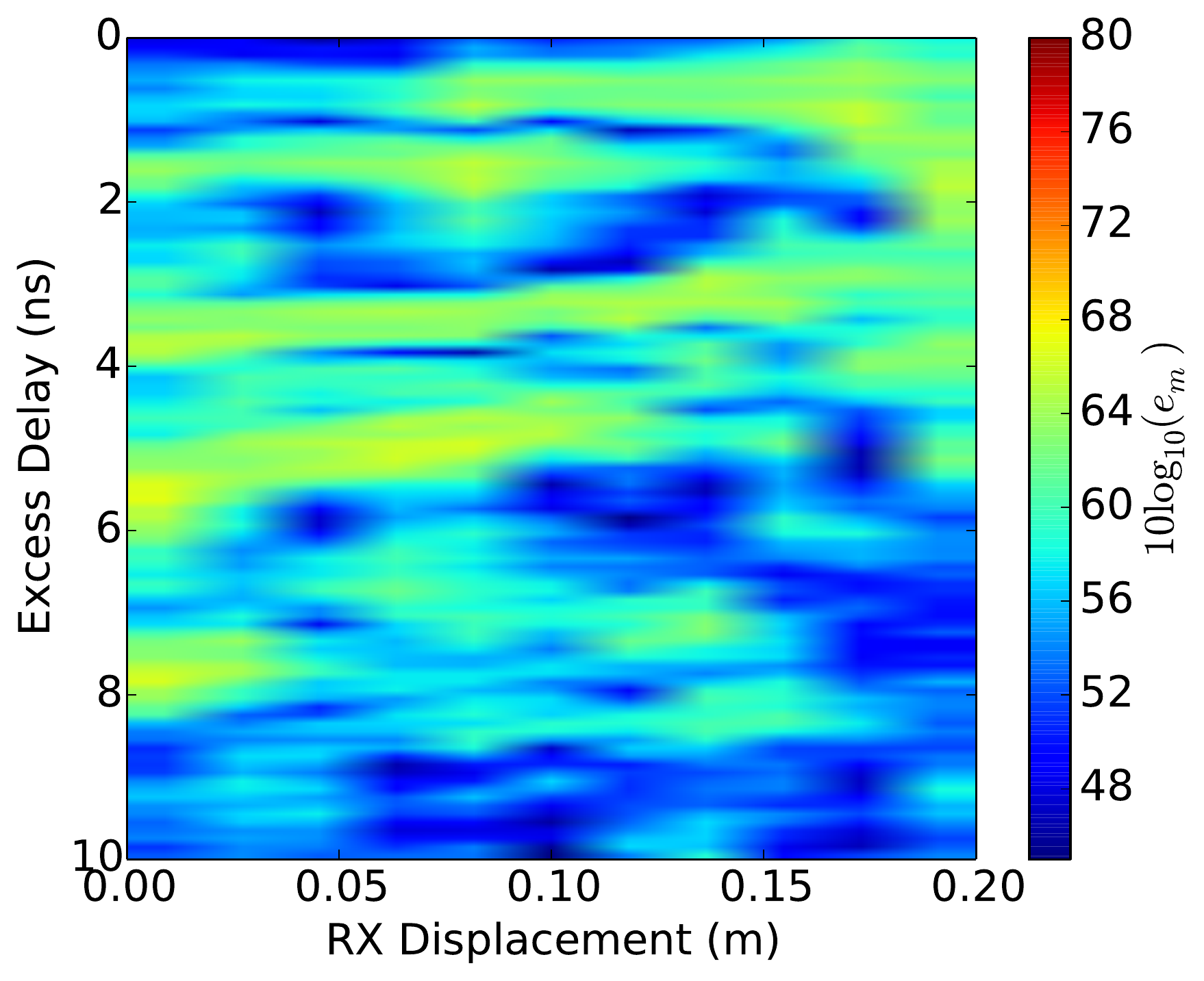}}
    \end{center}
    \caption{The energy observed in the CIR as a function of excess time delay and receiver displacement when (a) the room is empty and (b) there is a link line presence.}%
  \label{F:PDPplots}
\end{figure*}

We place our two UWB-IR transceivers in adjoining offices of an empty room as shown in Figures \ref{F:3240floorPlan} and \ref{F:radioOnStool} with antennas at 1.1 m height.  We measure $\log_{10}$ of the PDPs with time bin width $T_w=100$ ps. We measure PDPs (see Figure \ref{F:rxMove&empty}) at ten receiver positions spaced by 0.02 m while the room is empty.  Over the course of 0.20 m displacement, the energy in any given bin changes slowly due to small-scale multipath fading.  In particular, the changes in the first few nanoseconds have relatively slow changes.

Next, we run the same test, but with the person standing still on the link line in the middle of the empty room.  In this case, with the same settings and same displacement of the receiver, we see the PDPs shown in Figure \ref{F:rxMove&onLOS}.  The vertical scale in  Figure \ref{F:rxMove&onLOS} is identical to that of Figure \ref{F:rxMove&empty}.  One can see that the energy in the first few ns is dramatically smaller.

\begin{figure}[t!]
        \center{\includegraphics[width=3.5in]
        {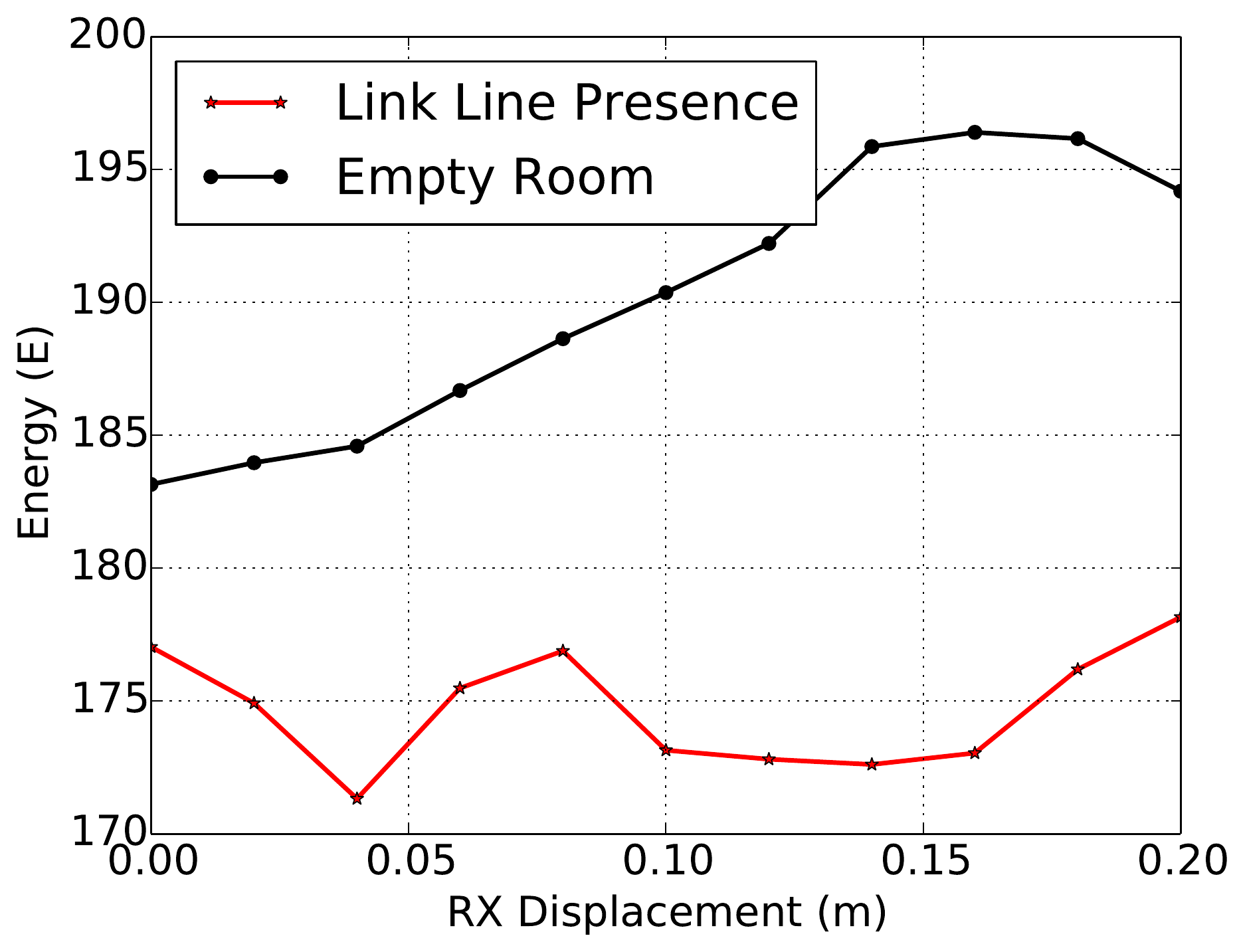}}
        \caption{\label{F:LOSpowerEmpty&LOS} The energy $E$ in the near-direct path as a function of receiver displacement for the empty-room case vs. link line presence.}
\end{figure}

We use a portion of the PDPs shown in Figure \ref{F:PDPplots} by plotting the energy in the first three nanoseconds. For simplicity, we define $E=10\log_{10}( e_0 )$ for the case when $T_w=3$ ns, that is, $E$ is $10\log_{10}$ of the energy in the first three nanoseconds of the impulse response measurement.  We plot $E$ as a function of receiver displacement for the two cases: empty-room vs. link line presence, in Figure \ref{F:LOSpowerEmpty&LOS}.

Comparing the empty-room and link line presence energy profiles, we observe that $E$ is much greater when the room is empty at any receiver displacement.  The difference in magnitude between the energies shows that we can use the energy in the first three nanoseconds of the CIR measurement as an indicator of link line presence even in the presence of small-scale fading.  It will later be shown through experimentation that we can use the measurements $E$ to detect link-line presence.  We note that for the remainder of this work, we use $E=10\logten (e_0)$ and $T_w=3$ ns.

%% file: RTI.tex
\subsection{Radio Tomographic Imaging} \label{S:RTI}
Radio tomographic imaging (RTI) has been shown to be an effective means for estimating the location of a person in the vicinity of an RF network that uses several stationary transceivers \cite{kaltiokallio2012grandma,kaltiokallio2012a,wilson10see}. It is also possible to apply RTI to measurements made with mobile transceivers, e.g., when the transceivers are attached to autonomous vehicles.


While several RTI methods have been developed, we apply shadowing-based RTI, which leverages the attenuation of RF transmissions moving people in order to infer their location. This form of RTI lends itself well to the UWB measurements at our disposal since we expect to be able to distinguish link line presence and no link line presence despite the effects of small-scale multipath fading.

We denote $E_l[n]$ to be the energy measurement $E$ for link $l$ at time $n$.  Then, the energy decrease between two measurements $n-1$ and $n$ is given as
\begin{equation} \label{E:energyChange}
\Delta E_l = E_l[n-1] - E_l[n].
\end{equation}
We note that a ``link'' is defined by the locations of the transmitter and receiver and refers to the two communicating transceivers.  Thus a link $l = (\mathbf{z}^{tx}, \mathbf{z}^{rx})$, where $\mathbf{z}^{tx}$ and $ \mathbf{z}^{rx}$ are the coordinates of the transmitter and receiver.  \emph{Other measurements made between any two devices at approximately the same coordinates are considered to be made on the same link}.  In the context of localization, when $\Delta E_l > 0$, we use this as evidence of link line presence and as input to radio tomographic imaging.

Let the change in energy on each link be formed into a measurement vector $\mathbf{y} = [\Delta E_1, \ldots, \Delta E_L]^T$. In order to generate an image, we assume that the total attenuation for each link is the sum of the attenuations caused by the voxels the link line passes through, i.e., 
\begin{equation} \label{S:weightedAttenuation}
y_l = \sum_{i=1}^M w_{l,i} x_i
\end{equation}
where $x_i$ represents the $i$th voxel in an image vector $\mathbf{x}$ containing $M$ voxels and $w_{l,i}$ is a weighting factor for quantifying the contribution of $x_i$ to the overall attenuation $y_l$ for link $l$. We use the weighting method
\begin{equation} 
  w_{l,i} = \begin{cases}
 \frac{1}{A_l} & \text{if } d^{tx}_{l,i} + d^{rx}_{l,i} < d_l + \lambda \\
 0 & \text{otherwise}
\end{cases}
\label{E:weightCalc} 
\end{equation}
\noindent where $d^{tx}_{l,i}$ and $d^{rx}_{l,i}$ are the distances between the centroid of voxel $i$ and the transmitter and receiver of link $l$, $d_l$ is the distance between the transmitter and receiver of link $l$, $A_l$ is the area of the ellipse, and $\lambda$ is the excess path length of the ellipse (used to control the width of the ellipse).

We can write the attenuations for all of the links in the network in matrix form as follows
\begin{equation} \label{E:MatrixEq}
\mathbf{y} = \mathbf{Wx} + \mathbf{n}
\end{equation} where $\mathbf{n}$ is the noise contribution and $\mathbf{W}$ is the $L \times M$ weighting matrix.  We solve for $\mathbf{x}$ using a regularized least-squares approach \cite{wilson09a,wilson10see,kaltiokallio2012a,kaltiokallio2012grandma}.

%% file: ExperimentalVerification.tex
\section{Experimental Verification} \label{S:Experiments}
 
In this section, we perform two measurement campaigns to: first, test that the changes in the channel due to small-scale multipath fading can be distinguished from differences in the channel measurement due to temporal fading; and second, test if we can perform DFL with mobile transceivers by localizing people based on the links whose channels show changes because of temporal fading.  

\subsection{Link Line Presence Detection Experiment} \label{S:LineCrossingExp}
\begin{figure}[t!]
        \center{\includegraphics[width=3.5in]
        {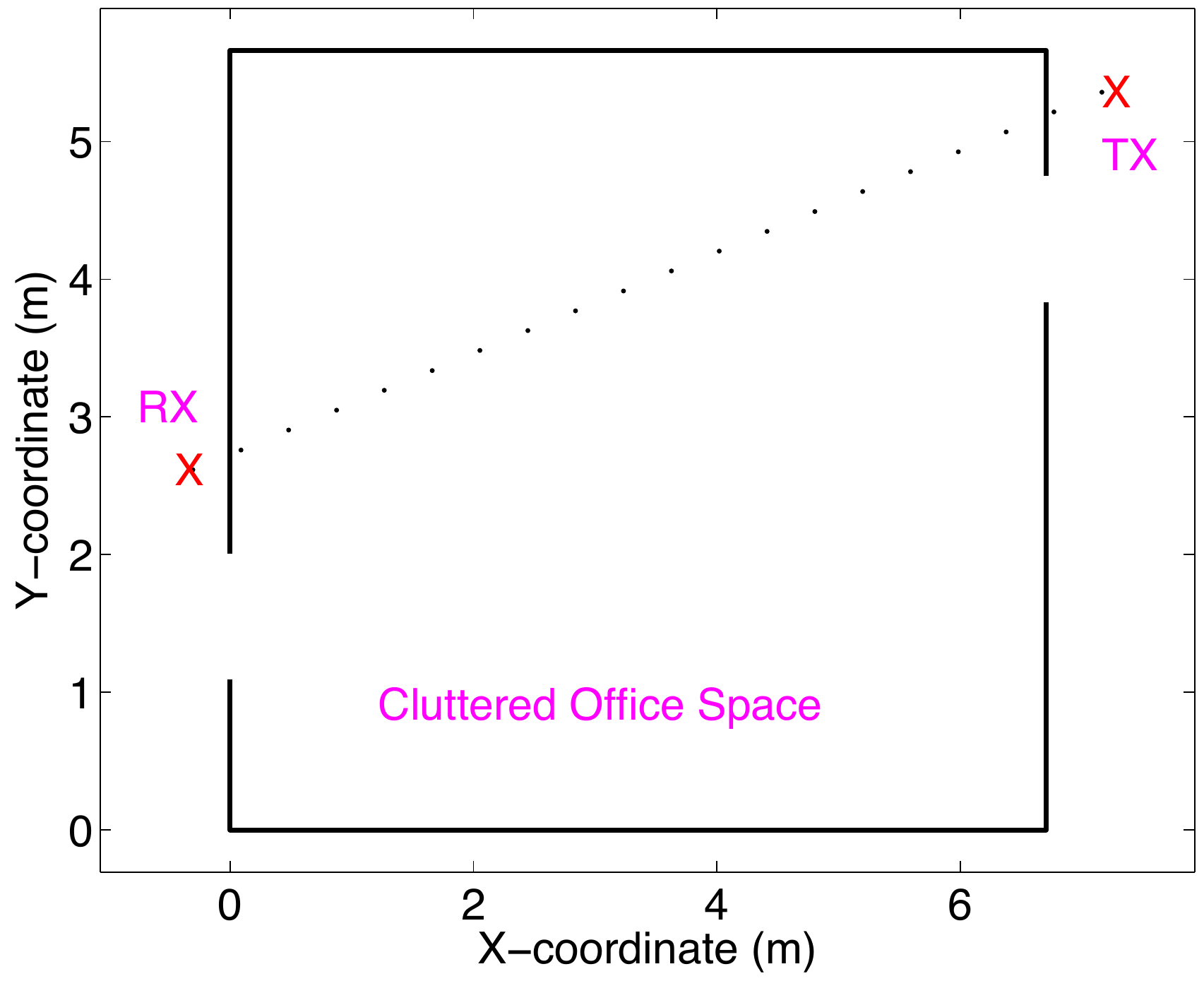}}
        \caption{\label{F:csGradLab}The floor plan of the office space used to perform link line crossing tests.  The office space is full of office furniture, creating a cluttered environment.}
\end{figure}

In the first measurement campaign, our goal is to detect when a person is on a link line formed by two wireless UWB transceivers while one of the transceivers is moving.  Being able to detect link line presence while one transceiver is moving demonstrates the feasibility to distinguish temporal fading from small-scale multipath fading.  We choose a cluttered room with a couch, several desks, bookcases, and chairs inside our engineering building  as the experimental site.  The layout of the room is shown in Figure \ref{F:csGradLab}.

\begin{figure}[t!] 
        \center{\includegraphics[width=3in]
        {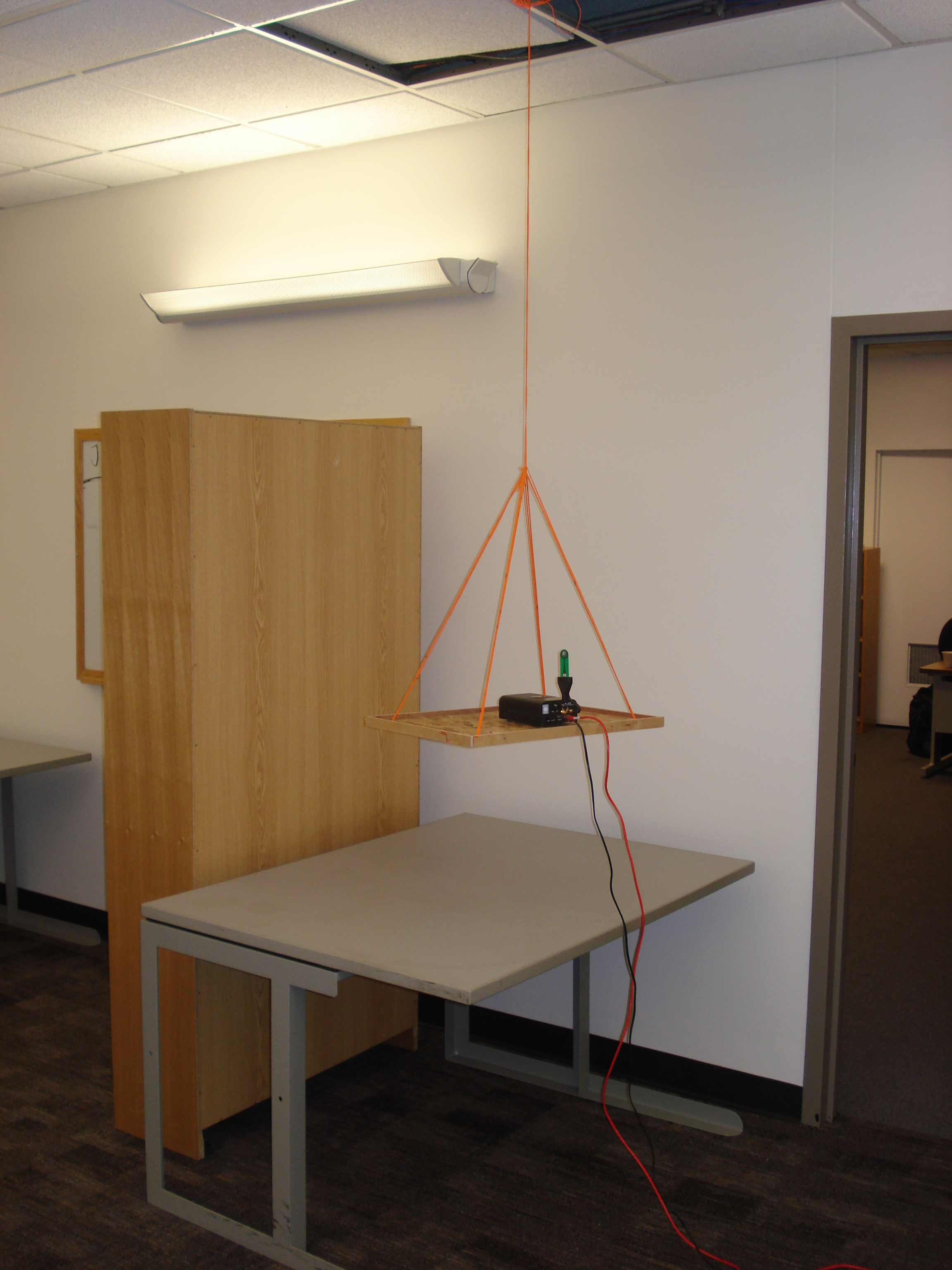}}
        \caption{\label{F:radioOnSwing}The receiver sits on a wooden platform which hangs from the ceiling.  The platform is equipped to swing parallel to the wall through 0.20 m.}
\end{figure}

We place two transceivers outside of the office space such that the link line is separated by two walls.  The receiver sits on a platform that is suspended from the ceiling (see Figure \ref{F:radioOnSwing}) while the transmitter is placed on a stool.
Both the transmitter and receiver are placed 1.1 m off the floor.  We attach strings to the hanging platform so that the receiver can be displaced by a short distance when the strings are pulled.

We choose to examine the effects of displacing the transmitter by 0.10 m and 0.20 m so that we could compare how a system would perform with different degrees of small-scale fading.  We perform two link line presence detection experiments using these two displacement distances.  In the first test, we move the platform repeatedly back and forth 0.10 m parallel to the wall. In the second test, we move the platform repeatedly back and forth 0.20 m parallel to the wall. In both cases, we move the platform back and forth every 2 seconds.  During the first test, a person walks at approximately 0.46 m/sec inside the office, crossing the link line 8 times (and a ninth time standing on the link line momentarily) at different points on the link line.  During the second test, a person again walks at approximately 0.46 m/sec inside the office crossing the link line 6 times.  The link line crossings are separated by at at least 10 seconds.  With a video recorder, we capture the time the person crosses the link line which we use to compare the measured and actual time of crossing.  Throughout the tests, the wireless channel is measured approximately 3 times per second. 

\subsection{Link Line Presence Detection Results} \label{S:LineCrossingResults}

\begin{figure}[t!] 
        \center{\includegraphics[width=3.5in]
        {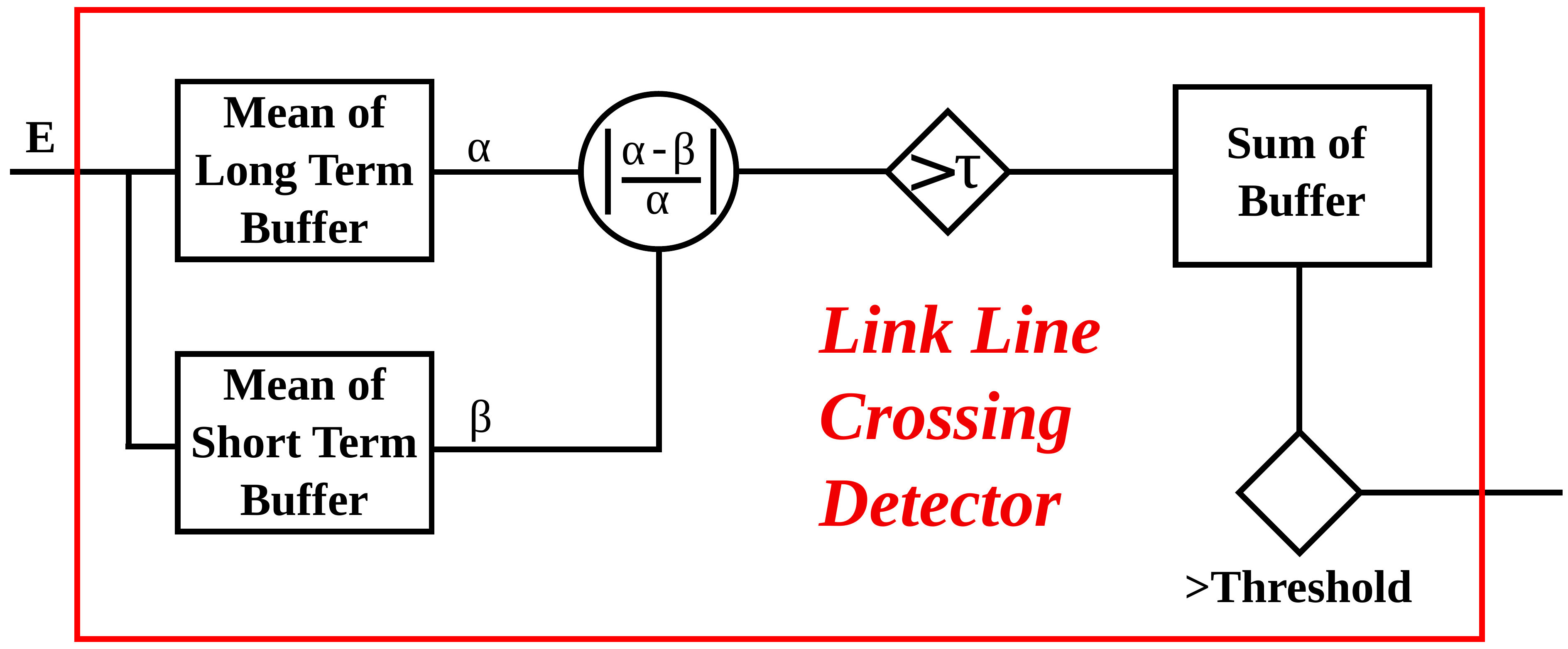}}
        \caption{\label{F:llcd} Moving average based detection \cite{youssef07}: Detector is 1 if the difference between short-term and long-term average exceeds $\tau$ multiple times during a time interval.}
\end{figure}

We found in Section~\ref{S:thruWall_MEB3240_results} that $E$ could be used to detect link line presence.  Although any link line presence detector could be used, we choose to implement the moving average based detector from Section from Section 4.3.1 in \cite{youssef07} because of its straightforward implementation and its accuracy in detecting link line crossings (see Fig. \ref{F:llcd}).
 The moving average based detector adds an $E$ measurement to a short and long term buffer.  The long term buffer stores the static behavior of the link while the short term buffer stores the current behavior.  Upon adding a new $E$ measurement to the buffers, the detector computes the relative difference between the means of the two buffers.  When the relative difference exceeds a threshold $\tau$, an event is detected.  These events are stored in a buffer that is summed with every new added event.  If the sum of the buffer exceeds another threshold, a link line presence is detected.  We let $\tau = 0.016$ and use the best parameters described in \cite{youssef07}.

\begin{figure}[t!] 
        \center{\includegraphics[width=3.5in]
        {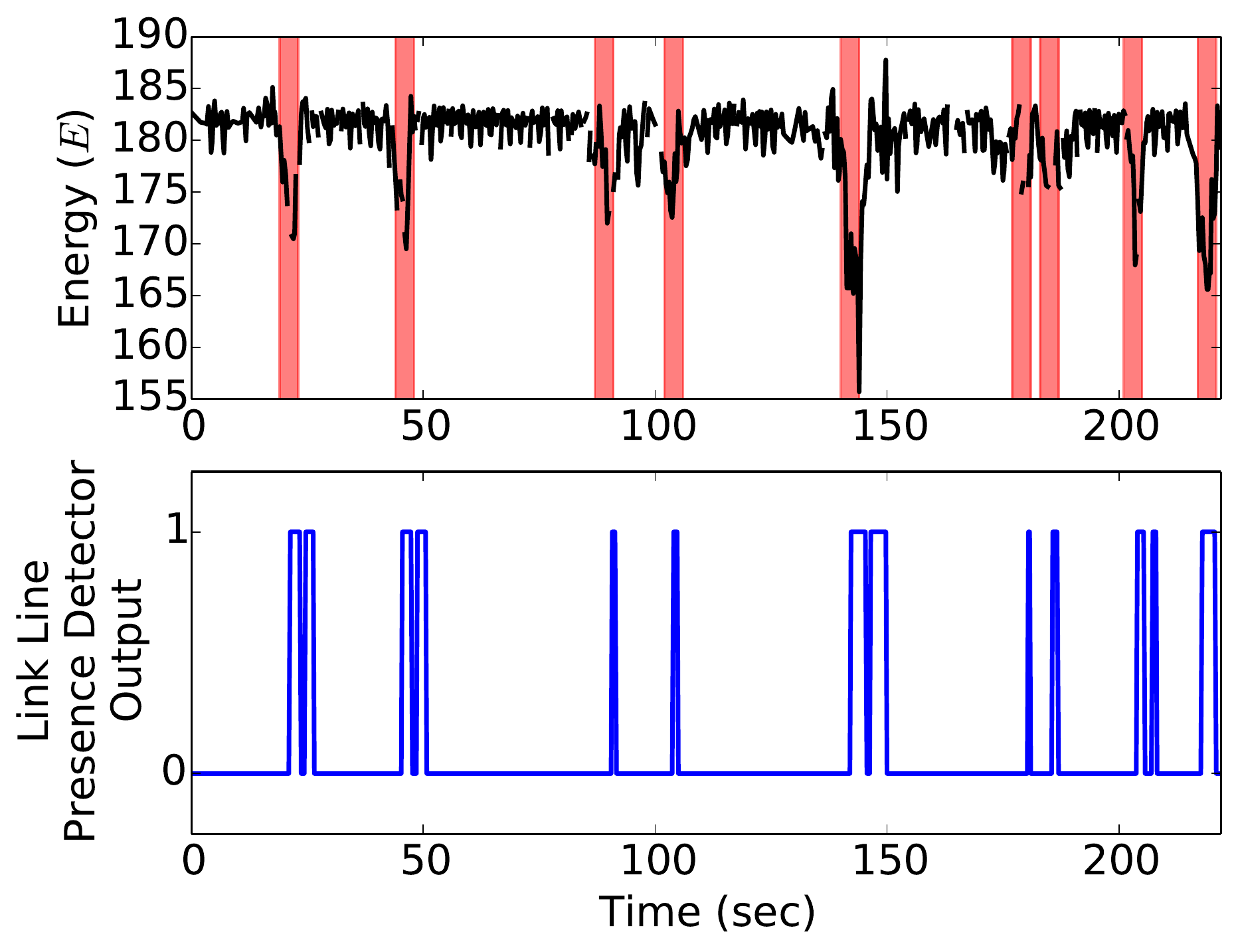}}
        \caption{\label{F:10cmScore}Measurements $E$ over time with the output of the link line presence detector when the UWB receiver position is displaced up to 0.10 m.}
\end{figure}

\begin{figure}[t!] 
        \center{\includegraphics[width=3.5in]
        {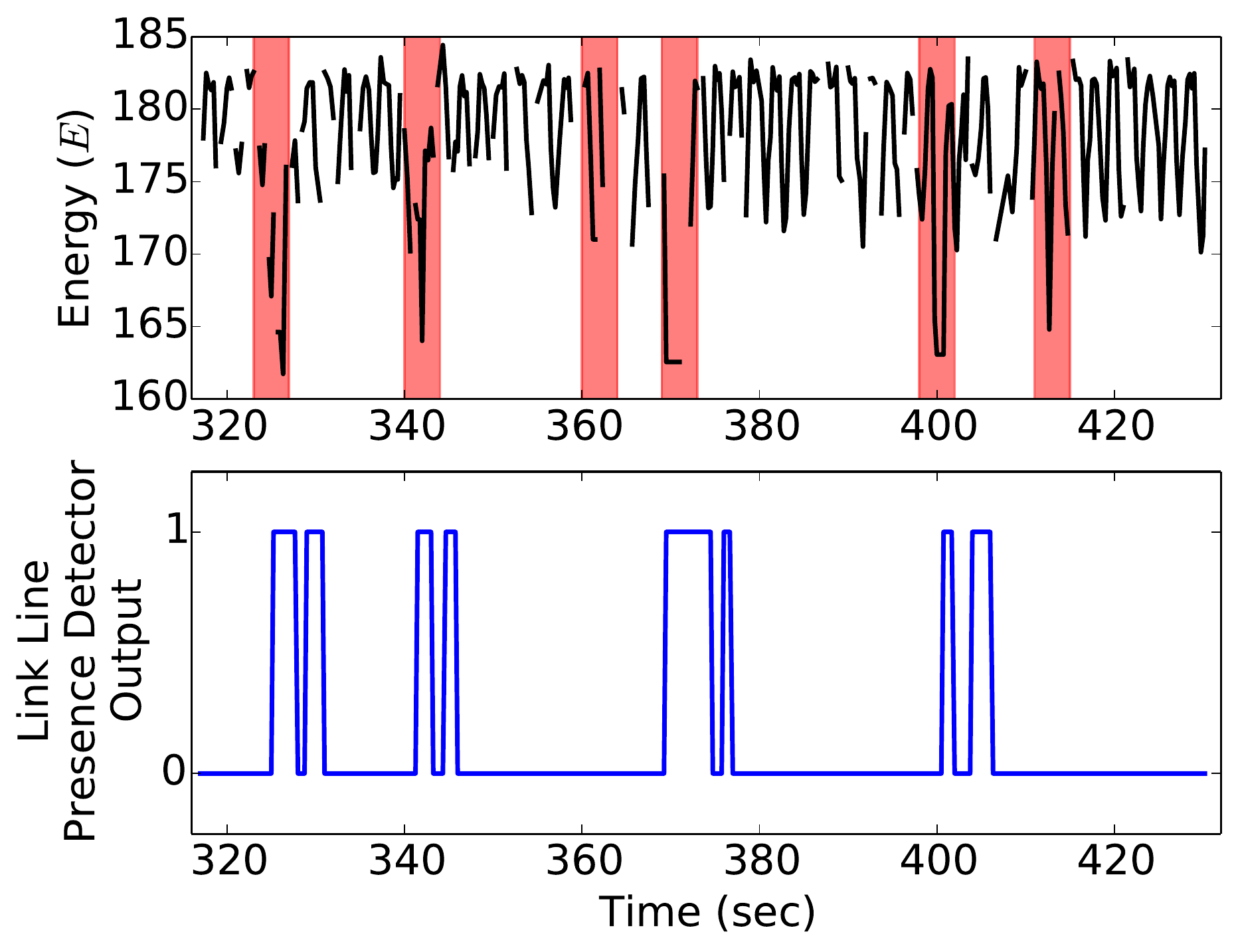}}
        \caption{\label{F:20cmScore}Measurements $E$ over time with the output of the link line presence detector when the UWB receiver position is displaced up to 0.20 m.}
\end{figure}

We use this detector with the measurements recorded during this experiment.  Figure \ref{F:10cmScore} and Figure \ref{F:20cmScore} show the results of the experiment.  
A link line presence is correctly detected when at least one link line crossing event occurs during the time the person is actually crossing the link line.  This is reasonable considering that a person is within 3 ns of excess delay for potentially several seconds as they cross the link line (see the measurements $E$ for the 140 second mark of Figure \ref{F:10cmScore}).  
Our algorithm accurately detects the nine occurrences of link line presence when the receiver is repeatedly moved through 0.10 m.  In contrast to the previous experiment, errors result when we move the receiver repeatedly through 0.20 m.  Of the six occurrences of link line presence, four link line presence events are accurately detected.

We observe two important facts from our two tests.  First, we show that detecting link line presence is more accurate when the transceiver is displaced by 0.10 m than when the transceiver is displaced 0.20 m.  The amount of small-scale multipath fading in the test with 0.20 m transceiver displacement is much greater than in the test with 0.10 m transceiver displacement.  Thus, in the test with 0.20 m transceiver displacement, there are instances when temporal fading could not be distinguished from small-scale multipath fading.  In addition, small-scale multipath fading can be severe enough to appear like the effects of temporal fading.  Second, we show that we can reliably distinguish small-scale multipath fading from temporal fading when the transceivers are displaced by no more than 0.10 m.  Clearly, there are alternative ways other than our algorithm to distinguish temporal fading from small-scale multipath fading, however, our results show the feasibility of accomplishing this task.

\subsection{Person Localization} \label{S:targetLocalization}
Knowing that we are able distinguish the effects of temporal fading and small-scale multipath fading, our goal in the second measurement campaign is to show that a system can perform DFL with mobile transceivers by collecting mobile UWB-IR measurements and computing a transceiver's relative coordinate locations to locate moving people based on the links whose channels show changes because of temporal fading. We describe how we adapt RTI for channel measurements made with mobile transceivers to create an image that shows the presence of a person inside a network.

\subsubsection{Experiment} \label{S:RTIexperiment}
\begin{figure}[t!] 
        \center{\includegraphics[width=3.5in]
        {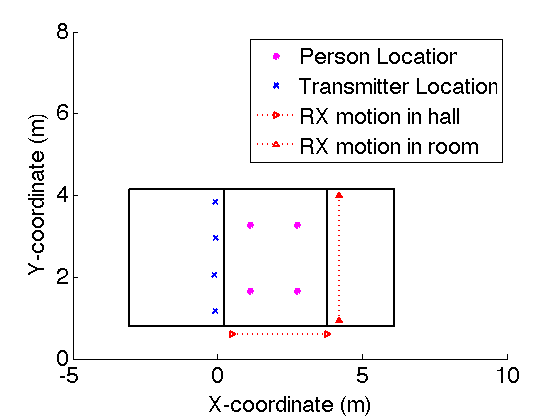}}
        \caption{\label{F:churchLayout} A room layout of the test site used to perform localization.}
\end{figure}

We use a classroom whose walls are made of brick for our experiment test site.  The classroom has a few tables and chairs making it a semi-cluttered environment.  Figure \ref{F:churchLayout} shows the layout of the experimental site.

In our experiment, we manually move one of the UWB transceivers to collect mobile channel measurements.  We select four transmitter positions and for each transmitter position, we manually move the receiver on a track at approximately 0.08 m/s along the walls of the adjacent room and hall as shown in Figure \ref{F:churchLayout}.  The receiver measures the channel approximately every 100 ms while two laser-range finders approximately every 300 ms log the relative position of the receiver with respect to a fixed coordinate system.  Because the channel measurements and the laser-range finder coordinates are sampled at different times, we linearly interpolate to match each channel measurement to a receiver coordinate.  We call the coordinate at which the receiver measures the channel a \emph{mobile receiver coordinate}.  We perform this process for when the room is empty and while a person stands at each of the 4 positions in Figure \ref{F:churchLayout}.

To create images using RTI, the transceivers must have static coordinates.  Since our receiver coordinates are mobile (i.e., two receiver coordinates, and therefore link line endpoints, are most likely not the same), we choose $R$ \emph{reference receiver} coordinates and we associate each mobile receiver coordinate to one reference receiver coordinate.

We now describe how we form the measurement vector $\mathbf{y}$ in (\ref{E:MatrixEq}).  We first define each transmitter location to be at one of four locations $s$ where $s \in \left\{{1,2,3,4}\right\}$.  When the receiver is at the $j$th mobile receiver coordinate and the transmitter is at location $s$, we compute the energy measurement $E$ of the channel and denote it $E_{s,j}$.  We denote the $j$th mobile receiver coordinate when the transmitter is at location $s$ as $\mathbf{z}^{mc}_{s,j}$; and we denote the $r$th reference receiver coordinate $\mathbf{z}^{rc}_r$.

We associate each $E_{s,j}$ to the nearest reference receiver by finding $r$ such that
\begin{equation} \label{E:binCIRmeasurements}
\argmin_r \| \mathbf{z}^{mc}_{s,j} - \mathbf{z}^{rc}_r  \|
\end{equation}
where $\| \cdot \|$ is the Euclidean distance. The energy measurement $E_{s,j}$ then belongs to the set $\mathcal{E}_{r,s}$ where $r$ is the $r$th reference receiver and $s$ is the transmitter location.  For simplicity, we shorten the notation of the set to $\mathcal{E}_{l}$ where $l$ is the link formed by reference receiver $r$ and the transmitter at position $s$.  Finally, we denote the median of the values in set $\mathcal{E}_{l}$ as $M_{l}$.

We redefine the components of the measurement vector $\mathbf{y}$ presented in (\ref{E:MatrixEq}) to take into account multiple measurements per link.  We first define the change in median energy $\Delta M_{l}$ to be
\begin{equation} \label{E:changeInMedianE}
\Delta M_l = M^{cal}_l - M^{occ}_l
\end{equation}
where $M^{cal}_l$ is computed from measurements when the room is empty and $M^{occ}_l$ is computed from measurements when a person is in the room.  In the event no median can be computed because $\mathcal{E}_{l} = \varnothing$, we set $\Delta M_l = 0$. We are interested in finding the change in median energy between empty-room case and occupied-room case.  Thus we define the measurement vector $\mathbf{y} = [y_1, \ldots, y_L]^T$ where $L$ is the number of links created by transmitter-reference receiver pairs and $y_l$ for link $l$ is
\begin{equation} 
  y_l = \begin{cases}
 \Delta M_l  & \text{if $\Delta M_l > 0$} \\
 0 & \text{otherwise}
\end{cases}
\label{E:newRTImsrtComponent} 
\end{equation}

\subsubsection{Mobile RF Network Results} \label{S:RTIresults}

In this section, we show the images we produce using RTI and the effectiveness in estimating the location of the person.  We choose the number of reference receivers $R=67$; 30 in the room and 37 in the hall.  We evenly space the reference receivers in each set along each wall, the hall wall and the room wall as shown in Figure~\ref{F:churchLayout}, such that the distance between any two reference receivers is less than 0.12 m.  This ensures that we collect sufficient measurements for each reference receiver but that the reference receivers are close enough to adjacent reference receivers to avoid the effects of small-scale fading for medium to large delays.  

Figure~\ref{F:RTI_images} shows the images we produce using RTI. In all of these figures, excluding Figure \ref{F:RTI_position0}, the image shows high intensity near the location of the person, which follows the intuition that the link lines that pass through the voxels covered by the person will experience the greatest change in energy.  Using the voxel with the greatest intensity as the estimate of the person's location, we accurately estimate the location of the person to within the errors, shown in Table \ref{T:RTIerrors}:

\begin{figure*}[ht!]
     \begin{center}
        \subfigure[0.50 m error when the person stands at $j=1$.]{%
            \label{F:RTI_position0}
            \includegraphics[width=0.4\textwidth]{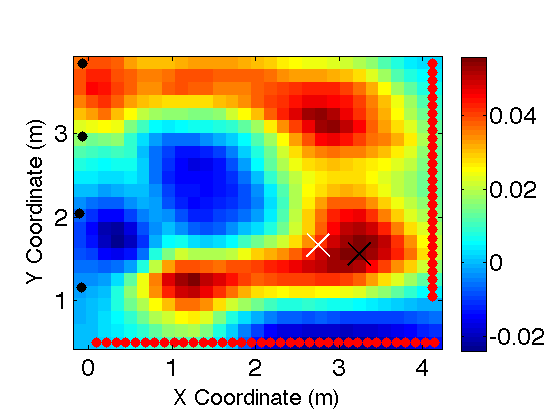}
        }%
        \subfigure[0.19 m error when the person stands at $j=2$.]{%
           \label{F:RTI_position1}
           \includegraphics[width=0.4\textwidth]{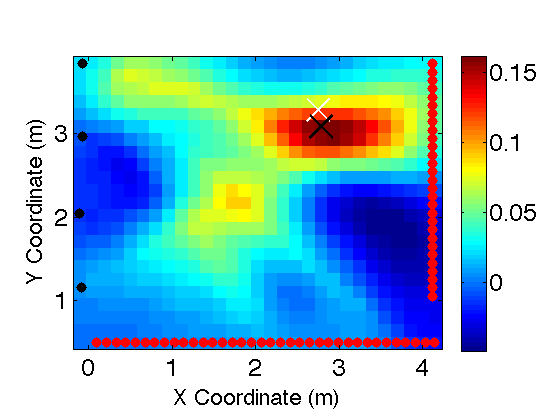}
        }\\ 
        \subfigure[0.13 m error when the person stands at $j=3$.]{%
            \label{F:RTI_position2}
            \includegraphics[width=0.4\textwidth]{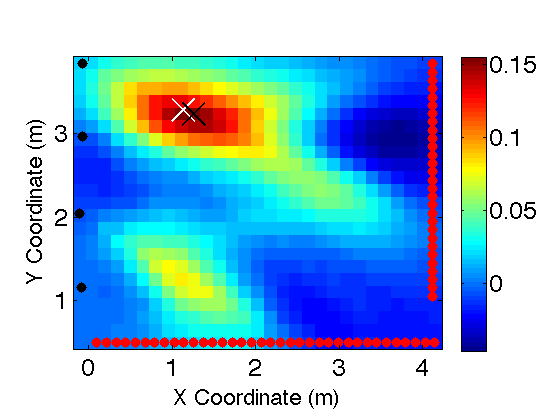}
        }%
        \subfigure[0.16 m error when the person stands at $j=4$.]{%
            \label{F:RTI_position3}
            \includegraphics[width=0.4\textwidth]{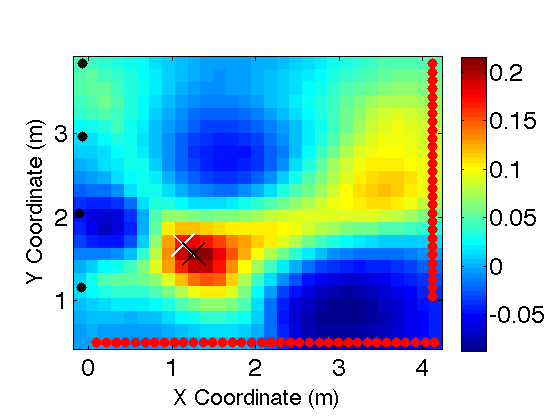}
        }%
    \end{center}
    \caption{%
        Images produced using RTI for the mobile transceiver experiment campaign.  The black X's are the true coordinates and the white X's are the estimated coordinates.  The black and red dots surrounding the image are the transmitter and reference receiver locations, respectively.  Areas in the image with higher intensity suggest a greater $\Delta M$ for link lines that pass through that area of the network.  
     }%
   \label{F:RTI_images}
\end{figure*}

\begin{table}[htb]
\caption{Error in Person's Estimated and Actual Location} 
\centering 
\begin{tabular}{c c} 
\hline\hline 
Standing Position & Error \\ [0.5ex] 
\hline 
1 & 0.50 m \\ 
2 & 0.19 m \\
3 & 0.13 m \\
4 & 0.16 m \\ [1ex] 
\hline 
\end{tabular}
\label{T:RTIerrors} 
\end{table}

The average of the four error values is approximately 0.25 m.  These results show strong evidence that we can reliably localize a person in a through-wall environment using few mobile transceivers.

%% file: RelatedWork.tex
%
\section{Related Work} \label{S:RelatedWork}
To the best of our knowledge, this paper presents a novel approach for using UWB-IR measurements for device-free or non-cooperative localization.  Traditional multistatic radar methods require that the UWB radios remain static \cite{chang04,reggiani09,rydstrom08,paolini08,mccracken2012hidden}.  These methods measure the excess delays of any new multipath components assumed to be due to a new person in the environment, from multiple pairs of UWB transceivers.  Each excess delay restricts the person to be within an ellipsoid of specified width, and assuming sufficient measurements, people's locations can be determined \cite{chang04,rydstrom08}.  However, the problem of finding one changed excess delay among dozens or hundreds of multipath components and noise is not at all a trivial problem \cite{mccracken2012hidden}.  Our methods are complementary by allowing the UWB radios to be mobile, and by monitoring for changes in the first few nanoseconds of excess delay to detect line presence.  These small excess delays are typically ignored in multistatic UWB \cite{paolini08}.  We then use the many channels measured by mobile devices to create a map of an area and show that accuracies of 0.25 m can be achieved.

This work is also different from other localization schemes that use narrowband wireless devices.  Localization of people inside a building has been demonstrated using calibration measurements on many link lines passing through the network \cite{viani2010electromagnetic,seifeldin2009nuzzer} as well as using RTI to image the location of people \cite{wilson09a,wilson10see,kaltiokallio2012a}.  In these studies, many static devices were need to achieve high accuracy localization.  In addition, narrowband devices provide one measurement for the entire channel.  We build upon RTI by showing that through-wall imaging can be performed with fewer devices that move around the perimeter of the area.

Our work focusses on the first few ns of a measured CIR to provide information about the absence or presence of a person, whereas in \cite{aftanas09}, an image of an environment is produced using the time-delays between a transmitted pulse and a reflected pulse off of an object in the environment.  The system is similar to synthetic aperture radar.  This work shows that in an otherwise empty room, large conducting objects (e.g. a metal slab, and a large wooden cabinet) located a meter away from the wall can be imaged.  Unlike our work, however, no experiments were performed with people.

%% file: Conclusion.tex
\section{Conclusion and Future Work} \label{S:Conclusion}

In this paper, we developed methods for device-free localization using a small number of transceivers in motion.  We accomplished this by measuring the energy of the first few nanoseconds of a received UWB signal and observing the changes between past and current measurements. Through several measurement campaigns, we showed that it is possible to detect the presence of a person in the environment when both small-scale fading and shadowing affect the received signal of wideband devices.  Our measurement campaigns demonstrated that when a receiver was kept to within a displacement of 0.20 m, we can accurately detect when a person is on a link line more than 80\% of the time.  In another campaign, we demonstrated that we can estimate the location of a person to within 0.25 m of their actual location.  These results were achieved using our mobile implementation of radio tomographic imaging.

We have used UWB radios as a means to measure energy in the first few nanoseconds of the received signal.  Future work includes localizing people in a network using 802.11 links as a means to gather measurements.  Additional work could explore alternate statistical methods to detect changes in the received signal.